\documentclass[10pt, conference, compsocconf]{IEEEtran}

\usepackage[utf8]{inputenc}
\usepackage[english]{babel}
\usepackage[dvipsnames]{xcolor}
\usepackage{cite}
\usepackage{graphicx, tikz, epstopdf, adjustbox, stfloats}
\usepackage{balance}
\usepackage[hyphens]{url}
\usepackage[inline]{enumitem}
\usepackage[hidelinks]{hyperref}
\usepackage{todonotes}

\graphicspath{{figs/}}

\usepackage[]{cleveref} 
\crefformat{section}{Section~#2#1#3}
\crefformat{figure}{Fig.~#2#1#3}

\usepackage[final]{microtype}
\def\baselinestretch{0.95}
\makeatletter
\newcommand\subparagraph{%
    \@startsection{subparagraph}{5}
    {\parindent}
    {3.25ex \@plus 1ex \@minus .2ex}
    {-1em}
    {\normalfont\normalsize\bfseries}}
\makeatother
\usepackage[compact]{titlesec}
\let\subparagraph\relax 
\titlespacing\section{0pt}{6pt plus 4pt minus 2pt}{2pt plus 2pt minus 2pt}
\titlespacing{\subsection}{0pt}{4pt plus 2pt minus 1pt}{2pt plus 1pt minus 1pt}
\titlespacing{\subsubsection}{0pt}{4pt plus 2pt minus 1pt}{2pt plus 1pt minus 1pt}
\usepackage{etoolbox}
\makeatletter
\patchcmd{\ttlh@hang}{\parindent\z@}{\parindent\z@\leavevmode}{}{}
\patchcmd{\ttlh@hang}{\noindent}{}{}{}
\makeatother

\newcommand*\circled[1]{\tikz[baseline=(char.base)]{
        \node[shape=circle,draw,inner sep=2pt, Maroon, fill=Maroon] (char)
               {\color{white}\scriptsize\textbf{#1}};}%
        }

\begin{document}
    \title{AVFI: Fault Injection for Autonomous Vehicles}
    \author{\IEEEauthorblockN{Saurabh Jha\IEEEauthorrefmark{4},
                             Subho S. Banerjee\IEEEauthorrefmark{4},
                             James Cyriac\IEEEauthorrefmark{2},
                             Zbigniew T. Kalbarczyk\IEEEauthorrefmark{2} and           
                             Ravishankar K. Iyer\IEEEauthorrefmark{2}\IEEEauthorrefmark{4}}
        \IEEEauthorblockA{\IEEEauthorrefmark{4}Department of Computer Science,
                         \IEEEauthorrefmark{2}Department of Electrical and Computer Engineering,\\
            University of Illinois at Urbana-Champaign, Urbana IL 61801, USA.}}
    \maketitle
    
     \begin{figure*}[!bp]
        \centering
        \vspace{-.7cm}
        \includegraphics[width=0.98\textwidth]{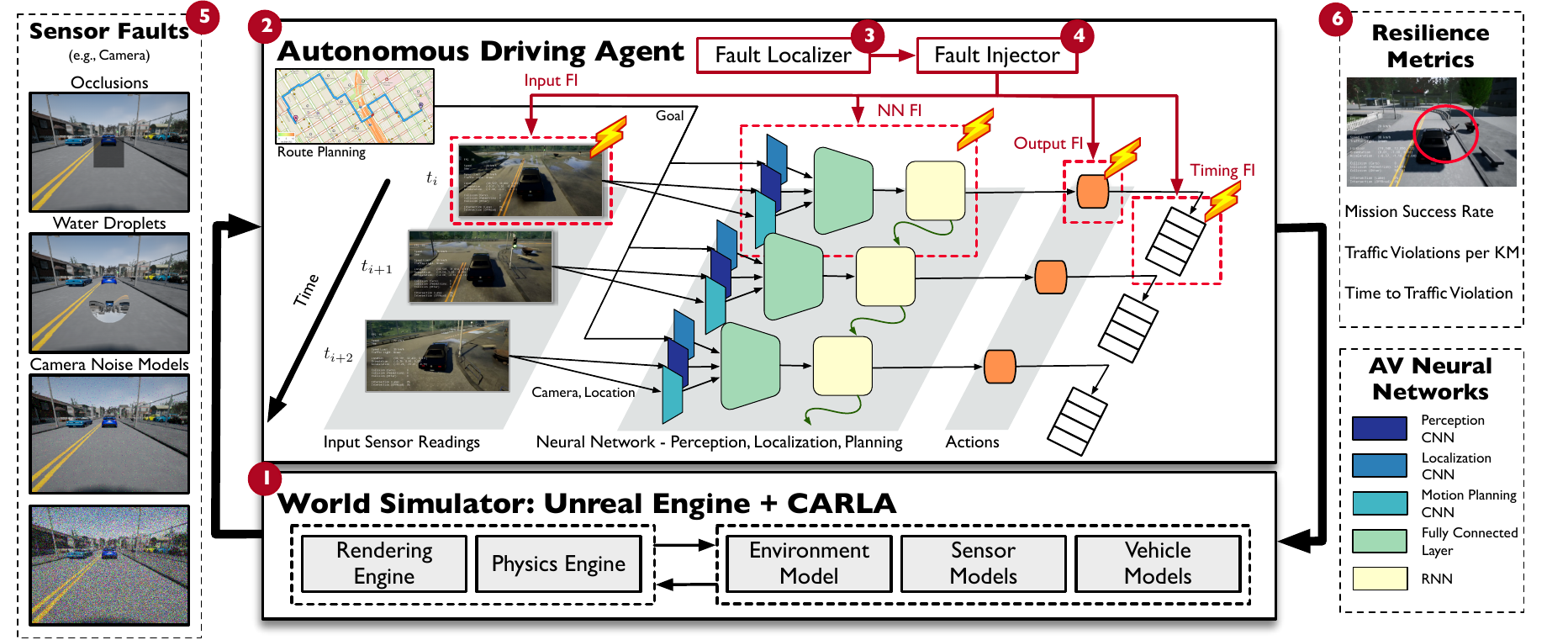}
        \vspace{-0.2cm}
        \caption{Overview of the AVFI approach: AVFI injects faults into sensor-compute-actuation systems of an autonomous vehicle.}
        \label{fig:overview}
    \end{figure*}
    
    \begin{figure*}[!bp]
        \centering
        \vspace{-.5cm}
        \begin{minipage}{.32\textwidth}
            \centering
            \includegraphics[width=\textwidth]{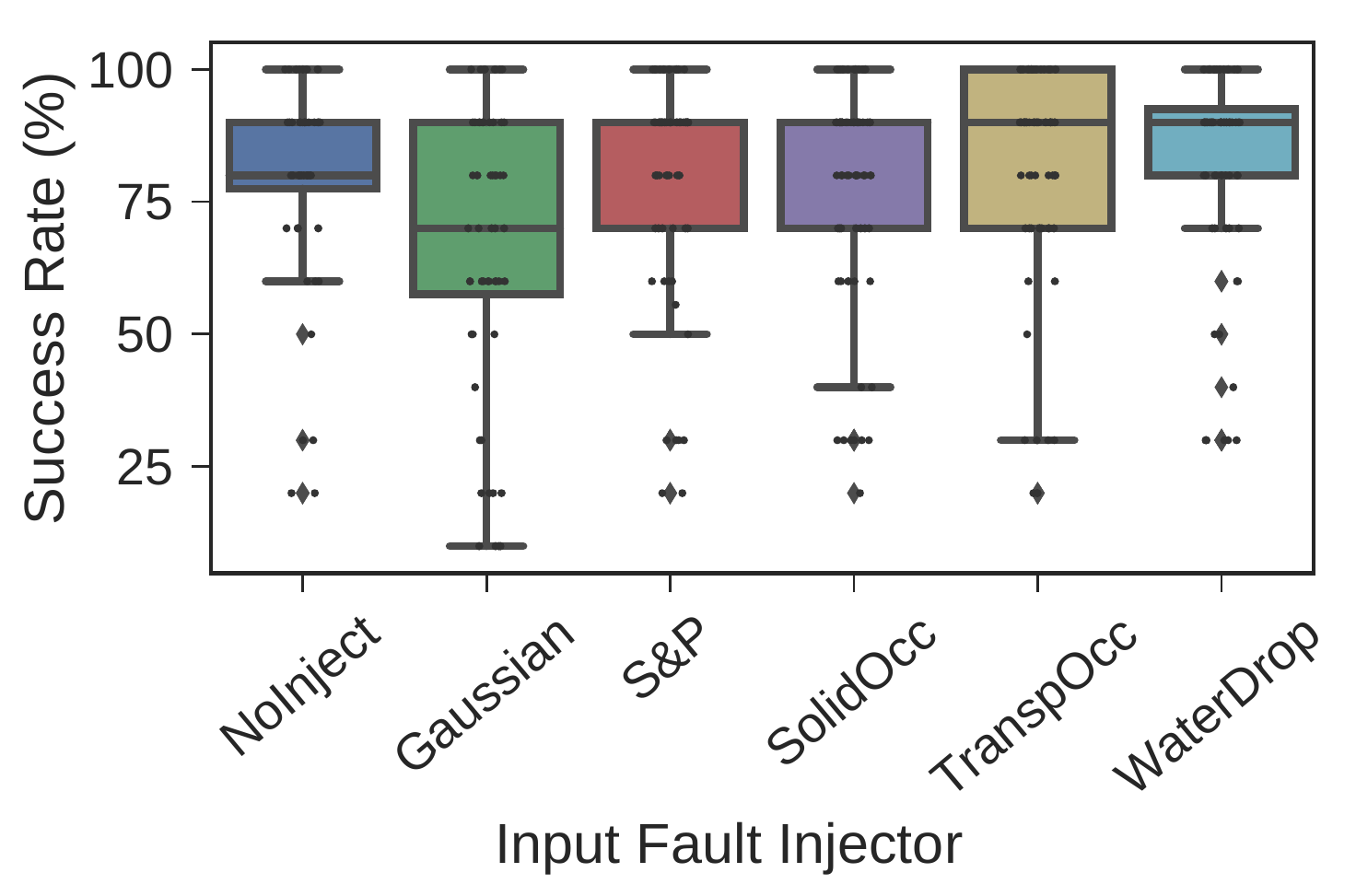}
            \vspace{-.7cm}
            \caption{Mission success rate for an autonomous vehicle with different input fault injectors.}
            \label{fig:success_rate}
        \end{minipage}%
        \hfill
        \begin{minipage}{.33\textwidth}
            \centering
            \includegraphics[width=\textwidth]{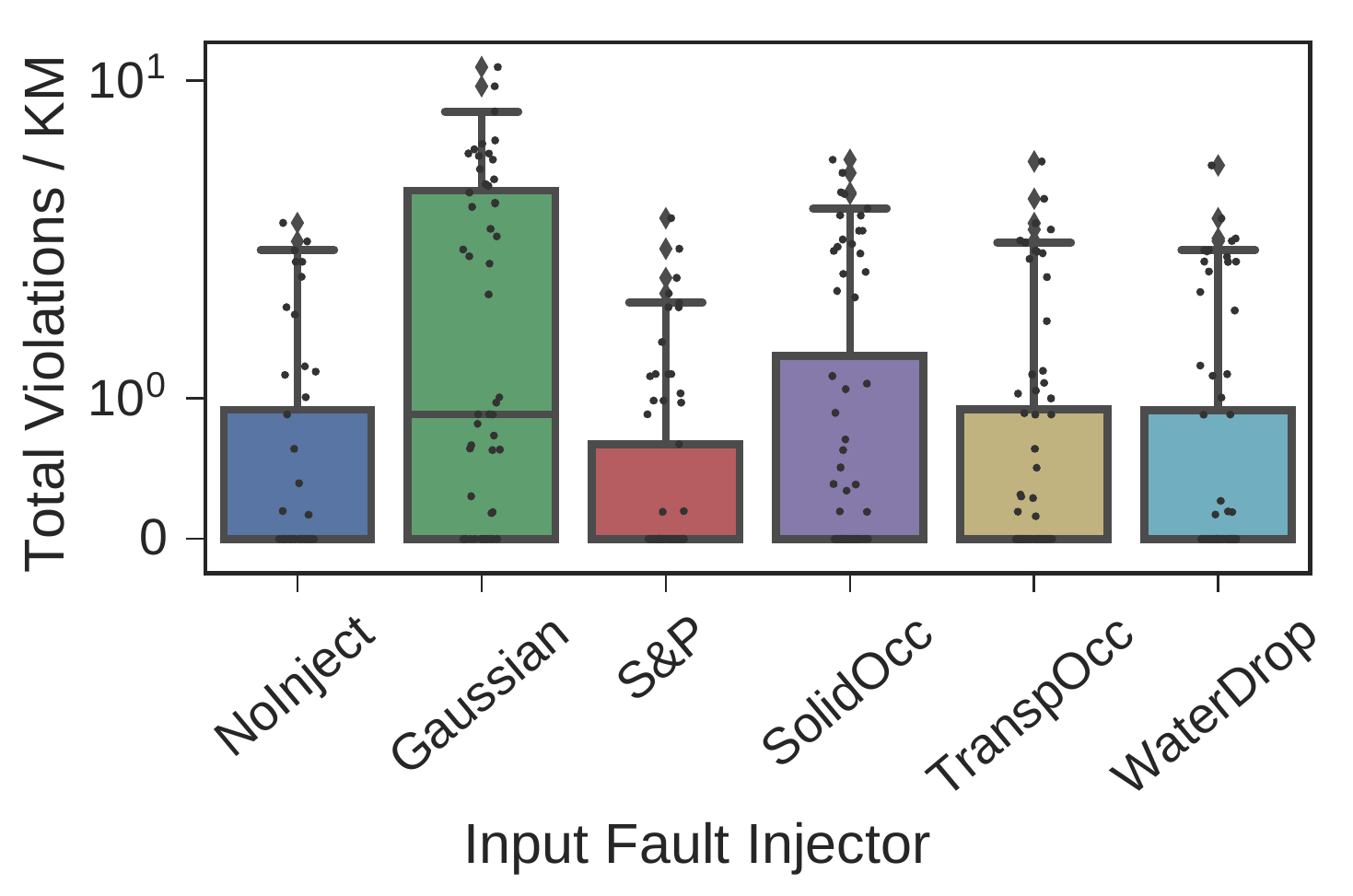}
            \vspace{-.7cm}
            \caption{Distribution of violations per km driven with different input fault injectors.}
            \label{fig:dpm_fi}
        \end{minipage}%
        \hfill
        \begin{minipage}{.32\textwidth}
            \centering
            \includegraphics[width=\textwidth]{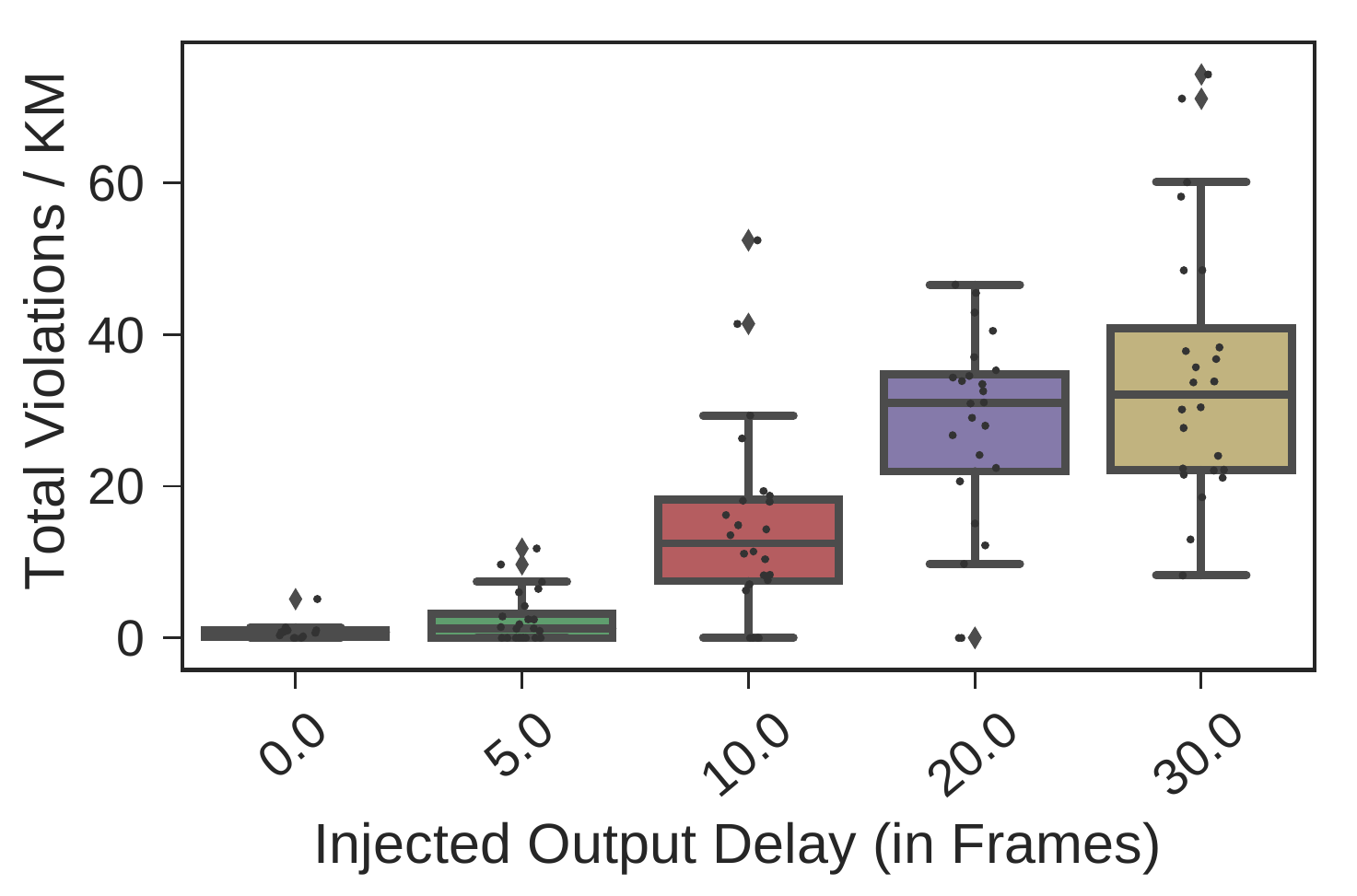}
            \vspace{-.65cm}
            \caption{Distribution of violations per km with increasing output delay between ADA and actuation.}
            \label{fig:dpm_delay}
        \end{minipage}%
        \vspace{-.6cm}
    \end{figure*}
    
    \section{Introduction} \label{sec:introduction}

Autonomous vehicle (AV) technology is rapidly becoming a reality on U.S. roads, offering the promise
of improvements in traffic management, safety, and the comfort and efficiency of vehicular travel.
With this increasing popularity and ubiquitous deployment, resilience has become a critical
requirement for public acceptance and adoption. Recent studies into the resilience of AVs have shown
that though the AV systems are improving over time, they have not reached human levels of
automation~\cite{DSN2018}. Prior work in this area has studied the safety and resilience of
individual components of the AV system (e.g., testing of neural networks powering the perception
function ~\cite{Li2017, Pei2017}). However, methods for holistic end-to-end resilience assessment of
AV systems are still non existent.

This paper presents AVFI (the Autonomous Vehicle Fault Injector), an important step towards constructing a methodology for end-to-end resilience
assessment of AV systems using fault injection. The tool
empirically validates the robustness of an AV system by introducing
faults to test AV resilience in situations that might otherwise be rarely tested. AVFI leverages a state-of-the-art AV simulation framework presented in
\cite{Dosovitskiy17}, and can perform fault injections in sensor inputs (e.g., following camera or
LIDAR fault models), in neural networks controlling the motion of the AV (e.g., to identify susceptibility to
random and adversarial noise in the training procedure), and in hardware/software components
(e.g., transient/permanent faults in processing fabric). The AVFI approach uniquely quantifies meaningful domain-specific failure metrics, e.g., number of traffic violations per kilometer driven, mission success rate and time to traffic violation. By using those metrics to evaluate safety, we demonstrate their comprehensive value. AVFI achieves those goals  so by simulating
real worlds, describing behavior of cars and pedestrians moving in that world, and evaluating resilience
metrics. Overall, we believe that AVFI can positively
influence the development and holistic testing of AV systems.

Our preliminary results validate AVFI's ability to introduce faults that lead to traffic violations. Those results are supported by failure characterization studies of AVs in the real world~\cite{DSN2018}. Our findings reiterate
the need for experimentation and analysis of failure models and modes for AVs.

    \section{AVFI: Autonomous Vehicle Fault Injector}\label{s:smfm}
\vspace{-0.15cm}
Our approach (see \cref{fig:overview}) uses (a) CARLA~\cite{Dosovitskiy17} as an urban driving simulator,  (b) the approach described in \cite{Codevilla2018} as an ADA (Autonomous Driving Agent) to control the AV and (c) AVFI as fault injection-based assessment engine for the ADA. AVFI has inbuilt fault models and provides methods for statistical
analysis of traffic violations.

\textbf{Autonomous Driving Agent \& World Simulator.} CARLA is an open
urban driving simulator (\circled{1}\footnote{\circled{*} refers to annotations in
\cref{fig:overview}.}). CARLA operates by running two components, the server and the client. The
server is responsible for generating the virtual urban environments, and the client functions as an
ADA. The server leverages Unreal Engine (popularly used in video games) as its rendering and physics engine. CARLA has an inbuilt
library of urban layouts, buildings, pedestrians, vehicles, and weather conditions (e.g., sunny,
rainy, and foggy) that can be used to simulate an urban environment. Further, it provides a variety
of sensors (e.g., camera, GPS, LIDAR) to use in AV simulations.

The ADA uses the approach described in \cite{Codevilla2018} as the controller; that in turn, uses an imitation learning-based convolution neural network (IL-CNN) for perception, planning, and localization (\circled{2}). In our test environment, the client is fed from a forward-facing RGB
camera sensor on the hood of the AV. The server sends sensor data, along with other
measurements of the car (e.g., speed, location) to the client. The controller is responsible for perception of sensor inputs and for producing an action that
describes the behavior of the AV. Its decisions are then sent from the client to the
server, which applies those commands to the AV’s actuators. In that way, an AV can complete missions,
i.e., navigating between way points in the simulated world. 

\textbf{Fault Models and Injector.}
AVFI runs fault injection campaigns in two steps: (a) selecting the location of faults (\circled{3})
(e.g., choosing specific neurons and layers in the IL-CNN) and (b) injecting the faults into the chosen
locations using the fault models mentioned below (\circled{4}). Broadly, AVFI can inject the following 
four classes of faults into the ADA.
\begin{itemize}[noitemsep,nolistsep,leftmargin=*]
    \item \emph{Data Faults}: AVFI injects data faults by manipulating sensor measurements (such 
    as camera images, LIDAR, and GPS) or world measurements (such as car spee or weather type) taken by the AV system. In the real world, sensor inputs can change because of (a) faulty sensors, 
    (b) changes in the external environment (such as fog or rain), and (c) unseen perturbations of images 
    (such as broken road sign posts). For example, AVFI intercepts the RGB camera sensor data from 
    the server, modifies the image according to a sensor-specific fault model (\circled{5}), and then 
    forwards it to the IL-CNN.
    
    \item \emph{Hardware Faults}: AVFI injects hardware faults by injecting single-bit, multiple-bit, and stuck-at faults in the hardware components of the autonomous systems, such as 
    processors, sensors, software, and communication networks. For example, AVFI can intercept and 
    corrupt a control command from the IL-CNN and then forward it to the server.
    
    \item \emph{Timing Faults}: AVFI injects timing faults into the communication paths of the 
    network, resulting in (a) delays in flow of data from one component of the AV system 
    to another, (b) loss of data, or (c) out-of-order delivery of the data packets. For example, 
    AVFI pauses the output of IL-CNN for $k$ frames and either replays or drops the outputs.
    
    \item \emph{Machine Learning Faults}: Errors in the machine learning models (such as neural 
    networks) during training or at runtime will lead to prediction errors. AVFI injects faults into 
    the neural network by adding noise into the parameters of the machine learning model (e.g., 
    weights of the neural network), which is modeled on real-world hardware failures. 
\end{itemize}

\textbf{Resilience Assessment.}
AVFI reports the following resilience metrics (\circled{6}).
\begin{itemize}[noitemsep,nolistsep,leftmargin=*]
    \item \emph{Mission Success Rate (MSR)} is the percentage of times that the autonomous agent was able to 
    complete a navigation mission in a fixed amount of time. Higher MSR values are representative of higher
    resiliency.
    \item \emph{Traffic Violations Per KM (VPK)} is the number of traffic violations (including lane 
    violations, driving on the curb, and collisions with pedestrians, cars, and other objects on 
    the streets) per kilometer driven in a fault injection campaign. Lower VPK values are representative of
    higher resiliency.
    \item \emph{Accidents Per KM (APK)} is the number of accidents (i.e., collisions with pedestrians/cars/etc.) per 
    kilometer driven in a fault injection campaign.
    \item \emph{Time to Traffic Violation (TTV)} is the time between a fault injection and its 
    manifestation as a traffic violation. Higher values of TTV imply that the system has more time to detect
    and correct its state to avoid traffic violations.
\end{itemize}
\vspace{-0.1cm}
    \section{Preliminary Results} \label{s:results}
\vspace{-0.1cm}
Initial experiments using AVFI on the IL-CNN-based ADA presented in
\cite{Dosovitskiy17} have shown promising results and point to the need for further
experimentation and analysis of failure models and modes for deep-learning-based ADA.

\cref{fig:success_rate} shows the increase in variance of the mission success rate with varying
sensor fault models across multiple test scenarios. \cref{fig:dpm_fi} shows a similar increase in
variability of traffic violations per km driven across a range of sensor fault injectors. The variability suggests that the overall decrease in success rate is correlated to the increase in traffic
violations per km. \cref{fig:dpm_delay} shows a significant increase in the number of traffic
violations per km  with the introduction of delays between the generation of output from the
agent's neural network and its actuation in the world model (i.e., using AVFI's timing fault
injector). Our simulation environment is configured to run at 15 frames per second; hence, a delay of
30 frames in \cref{fig:dpm_delay} corresponds to an overall delay of a mere 2 s between decision and
actuation. Our results and those in \cite{DSN2018} (e.g., data from Nissan), indicate a need to explore the real-time nature and constraints
associated with the AV.

    {
        \nocite{*}
        \scriptsize
        \balance
        \bibliographystyle{IEEEtran}
        \bibliography{IEEEabrv,references}
    }

\end{document}